\documentstyle[emulateapj,apjfonts,psfig]{article}

\lefthead{Salvesen et al.}
\righthead{3C~287}

\received{Date}
\revised{Date}
\accepted{Date}

\journalid{vol}{date}
\articleid{1}{4}
\paperid{id}

\cpright{AAS}{1999}
\ccc{x}

\begin{document}

\title{A Deep XMM-Newton Observation of the Quasar 3C~287}

\author{G.~Salvesen\altaffilmark{1}, J.~M.~Miller\altaffilmark{1},
E.~Cackett\altaffilmark{1,2}, \& A.~Siemiginowska\altaffilmark{3}}

\altaffiltext{1}{Department of Astronomy, The University of Michigan,
500 Church Street, Ann Arbor MI, 48109-1042; salvesen@umich.edu,
jonmm@umich.edu}
\altaffiltext{2}{Chandra Fellow}
\altaffiltext{3}{Harvard-Smithsonian Center for Astrophysics, 60
	Garden Street, Cambridge, MA 02138, aneta@head.cfa.harvard.edu}

       
\authoremail{salvesen@umich.edu}

\label{firstpage}

\begin{abstract}

We report on an {\it XMM-Newton} observation of the $z=1.055$ quasar
and Giga-hertz Peaked Spectrum (GPS) source 3C 287.  Our 62.3~ksec
observation provides an exceptional X-ray view of a prominent member
of this important subclass of active galactic nuclei (AGN).  The X-ray
spectra of 3C 287 are consistent with a simple absorbed power-law with
a spectral index of $\Gamma=1.72\pm 0.02$.  Our fits imply a bolometric
luminosity of $L = 5.8\pm 0.2 \times 10^{45}~{\rm erg}~ {\rm
s}^{-1}$ over the 0.3--10.0 keV band; this gives a mass lower limit of
$M_{BH min} \geq 4.6 \times 10^{7}~{\rm M_{\odot}}$ assuming X-rays contribute
10\% of the bolometric luminosity and radiation at the Eddington limit.  Iron
emission lines are common in the X-ray spectra of many AGN, but the
observed spectra appear to rule-out strong emission lines in 3C~287.
The simple power-law spectrum and absence of strong emission
lines may support a picture where our line of sight intersects a
relativistic jet.  Milliarcsecond radio imaging of 3C 287
appears to support this interpretation.  We discuss our results in the
context of different AGN sub-classes and the possibility that GPS
sources harbor newly-formed black hole jets.
\smallskip

\end{abstract}

\section{Introduction}

Connections between accretion disks and relativistic jets in black
hole systems are anticipated theoretically, but observational
constraints on this coupling remain elusive.  Some progress has been
made by tracking X-ray and radio flux variations in stellar-mass black
holes, for instance, where transient outbursts evolve over weeks and months
(e.g. Gallo, Fender, \& Pooley 2003).  However, observations of
stellar-mass black holes average over vastly more dynamical timescales and
much larger regions than observations of the supermassive black holes
powering AGN.  In a sense, every observation is an average, and every
image is coarse.  

Giga-hertz Peaked Spectrum (GPS) sources and Compact Steep-Spectrum (CSS) 
sources are well-known AGN subclasses.  Beyond the flux-based criteria used
to separate these classes, surveys reveal an important size difference:
GPS sources are generally less than 1 kpc in projected linear size, while
CSS sources can be 20 times larger (see, e.g., Edwards \& Tingay 2004).
These small size scales imply relatively
young jets from new activity.  Compact Symmetric Objects (CSOs) -- a
subclass of GPS and CSS sources -- represent an even more interesting
regime in which to explore disk--jet connections.  Proper motion
studies of some CSO sources imply jets that have only become active
within the last $10^{3}$ years (Conway 2002).

3C 287 is a well-known radio--loud quasar at $z=1.055$ that falls into the GPS
category, and perhaps also into the CSO category (Kellerman et al.\
1998).  Our preliminary analysis of a prior {\it Chandra} observation
suggested the presence of a broad, relativistic iron disk line.  Such
lines are excellent probes of the inner disk (e.g. Miller 2007),
making 3C 287 an even more exciting target in which to explore
accretion flows and to study disk--jet connections.  GPS quasars do not
necessarily have the same X-ray properties as GPS galaxies.  GPS galaxies
show large X-ray absoption columns while GPS quasars show significantly less 
absorption (Siemiginowska et al. 2008).  X-ray absorption in GPS galaxies
may result from an orientation coincident with an obscuring "torus",
possibly blocking the central engine from direct observation (Guainazzi
et al. 2006).  Although a modest survey of GPS sources has been performed
with Chandra (Siemiginowska 2003, 2008), deep X-ray observations of these 
sources are rare.

In the following sections, we report the results of fits to the
spectra of 3C 287 as measured using the {\it XMM-Newton} EPIC/pn,
MOS-1, MOS-2, and OM cameras.  Fits to these spectra suggest power-law
behavior typical of CSS sources and the absence of strong Fe K emission
lines.  We discuss these results as well as their implications on
emission mechanisms and our line of sight to 3C 287.

\section{Observations and Data Reduction}

{\it XMM-Newton} observed 3C 287 for a total of 62.3 ksec, starting on
2006 July 19 at 01:57:09 (UT).  The EPIC/pn and MOS cameras were run
in "PrimeFullWindow" mode with the "medium" optical blocking filter.
The RGS was run in the standard "spectroscopy" mode.  The OM obtained
a sequence of images in V (corresponding to UV in the source frame) in
order to sample any light from an accretion disk.  The results
reported in this analysis are restricted to the EPIC cameras and to
the OM; the RGS spectra are count-limited and do not add much
information.

SAS version 7.1 was used to reduce and filter the data and create
products from the raw ODF files.  The tasks "epproc", "emproc", and
"omichain" were used to create calibrated event lists from the pn, MOS, and OM
cameras.  Subsequent reduction was performed using "xmmselect".  The
pn event list was filtered to accept only good events (via
"PATTERN$\leq 4$" and "FLAG$=$0").  The MOS event lists were also
filtered to accept only good events (via "PATTERN$\leq 12$" and
"FLAG$=$0").  

\centerline{~\psfig{file=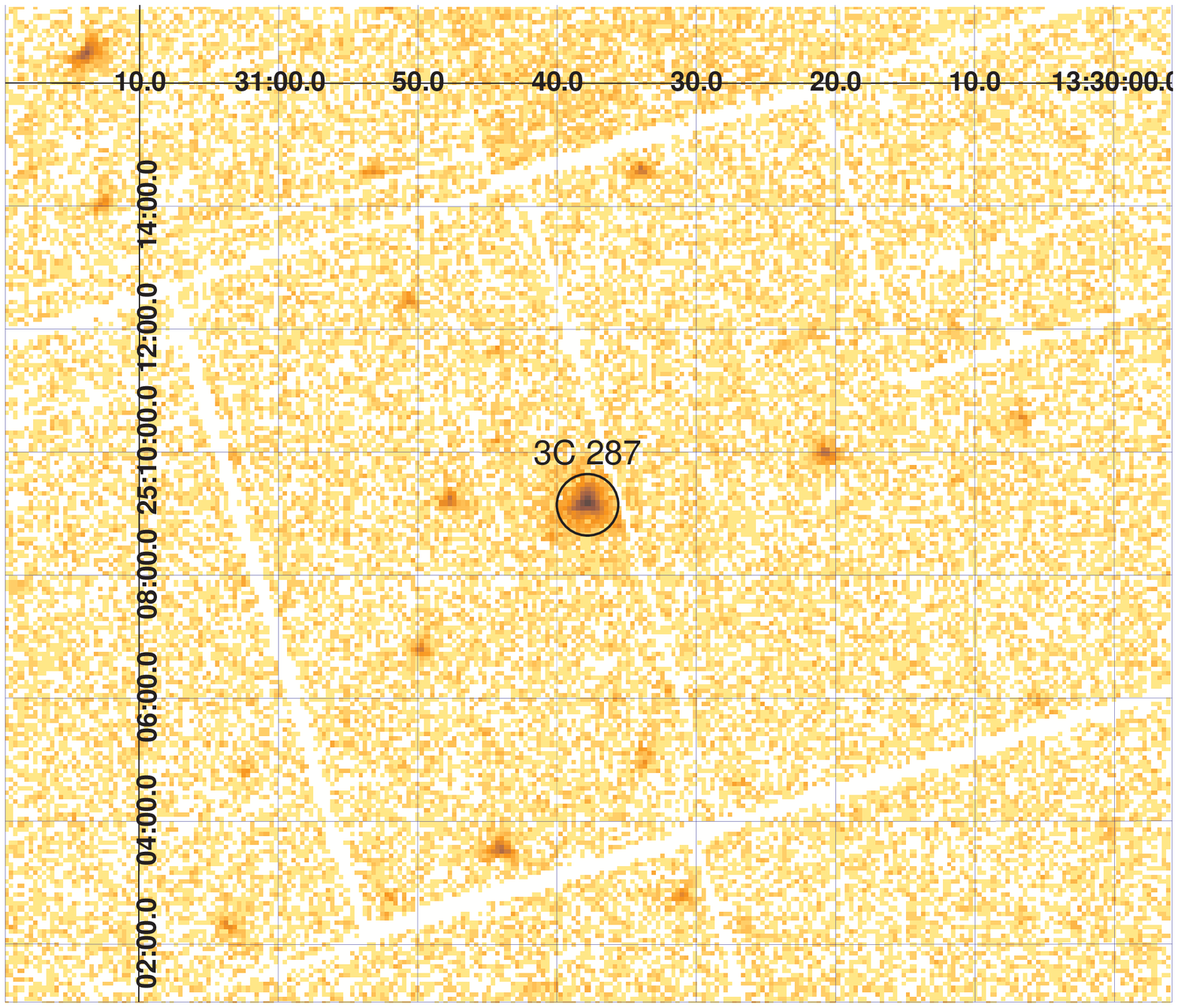,width=3.3in}~}
\figcaption[h]{\footnotesize The EPIC/MOS-2 X-ray image (0.3--10.0~keV)
of 3C 287 is shown above.  The source is circled.}
\medskip

Inspection of the EPIC lightcurves revealed strong flaring at the end
of the observation; therefore, only the first 48.1~ksec of the pn
exposure was used to create images and spectra, and only the first
53.8~ksec of the MOS exposures were used.  Circles with radii of 30
arcseconds centered on 3C~287 were used as source extraction regions.
Background regions of equivalent size were taken nearby to the source,
on the same CCD.  The pn source and background spectra were generated
by grouping channels 0--20,479 by a factor of 5; the MOS spectra were
generated by grouping channels 0--11,999 by a factor of 15.  Custom
response files for the EPIC spectra were generated using the tools
"rmfgen" and "arfgen".

To ensure valid constraints from the use of the $\chi^{2}$ fitting
statistic, the EPIC spectra were binned to require at least 20 counts
per bin using the FTOOL "grppha".  Spectra were then fit using XSPEC
version 11.3.2 (Arnaud 1996).  All EPIC spectra were fit in the
0.3--10.0~keV band as the cameras are best calibrated in this range.
All errors in this work are 90\% confidence errors as derived using
the XSPEC ``error'' command, unless otherwise noted.

\section{Analysis and Results}

\subsection{EPIC Analysis}

Model spectra were fitted to the data.  The pn and MOS spectral
data were fit jointly, allowing an overall normalization constant to float
between the spectra.  This normalization constant was within
3\% of unity.   All models were multiplied by the XSPEC model
"phabs" to account for photoelectric absorption by neutral hydrogen
along the line of sight, assuming default abundances.  The X-ray spectra
were jointly fitted using simple power-law models, where $\chi^{2}$ statistics
determined the goodness of each fit.  The three models we report are
spectral fits to a power-law, broken power-law, and line functions
incorporated into the power-law.  We added line functions to the power-law
model over a restricted energy range of 3.1--3.4~keV which is where
we expect to locate an Fe K emission line based on a redshift of $z =
1.055$ for 3C 287.  Spectral parameter values and fluxes, along with
the corresponding upper limits, are recorded for fits to
each of the models.  Investigation of our orientation relative to the
jet or disk comes from equivalent width methods, determined by the
upper limit of the normalization for the line 

\centerline{~\psfig{file=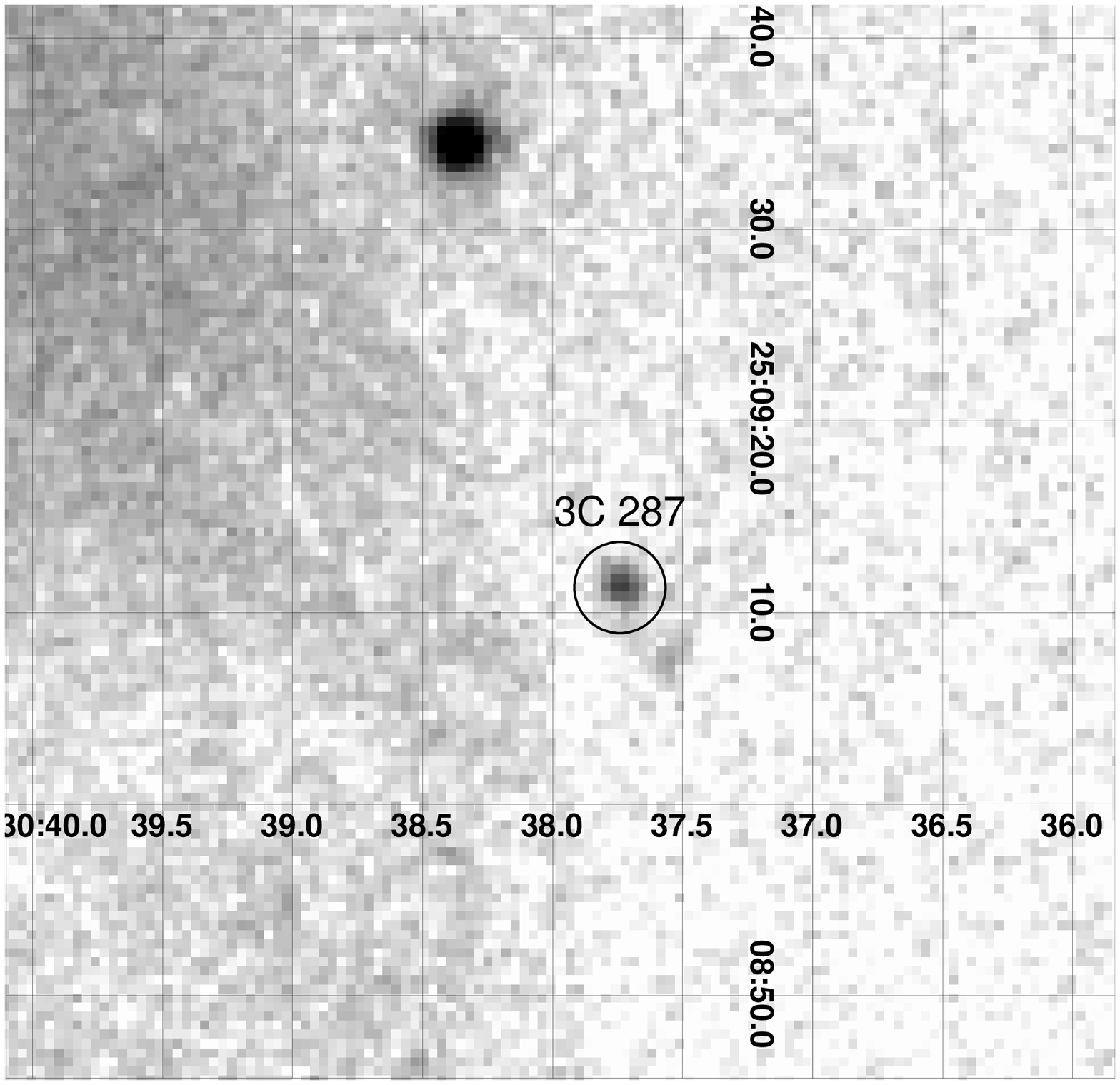,width=3.3in}~}
\figcaption[h]{\footnotesize The OM V-band image of 3C 287, generated from a 
single exposure, is shown above.  The source was found to have a V magnitude
of 18.2.  The apparent diffuse emission feature in the upper--left corner
is likely instrumental and has negligible impact on our OM measurements.}
\medskip

\noindent functions.
Both narrow and broad line fits were made to ensure robust limits
on the equivalent width.

We present images as observed by {\it XMM-Newton} and spectral fits along with 
parameters obtained using XSPEC.  Figures 1 and 2 show an X-ray and V-band
image of 3C 287 respectively, as observed by {\it XMM-Newton}.  The results
from the joint fits of MOS-1 and MOS-2 spectra to the pn spectra of 3C 287
over the 0.3--10.0~keV energy band are listed in Table 1.

Both a simple power-law model and power-law with a break in the 0.3--10.0~keV
range yield acceptable fits based on $\chi^{2}$ statistics.  The best-fit
power-law model gives $\chi^{2}/\nu = 650.3/598 = 1.09$, where $\nu$
is the number of degrees of freedom.  The best-fit broken power-law
model result is $\chi^2/\nu = 643.4/596 = 1.08$; this is not a
statistically significant improvement over the simple power-law model.
Spectral fits using a power-law model are shown in Figure 3.  The simple
power-law model gives a flux of $F = 7.1\pm 0.2 \times 10^{-13}$~ergs~
${\rm cm}^{-2} {\rm s}^{-1}$ over the 0.3--10.0~keV band.  We calculate a
corresponding bolometric luminosity of $L = 5.8\pm 0.2 \times
10^{45}~{\rm ergs}~ {\rm s}^{-1}$ based on the stated redshift $z =
1.055$, allowing us to place a minimum mass estimate on the central
engine.  The lower mass limit of 3C 287 is $M_{BH min} \geq 4.6 \times
10^{7}~ {\rm M_{\odot}}$, assuming that the X-ray luminosity accounts
for 10$\%$ of the bolometric luminosity and radiation at the Eddington
luminosity limit.  We calculate a recession velocity of $v_{rec} =
(0.62)c$ from the redshift, corresponding to a distance of $d = 2.6~
{\rm Gpc}$ assuming $H_0 = 71 ~{\rm km}~ {\rm s}^{-1}~ {\rm Mpc}^{-1}$.
This is the distance we assume when calculating the luminosity.

An interesting feature in the power-law model joint fits is the
absence of a strong Fe K emission line centered at a rest frame energy of
6.4~keV.  Figure 3 plots spectra in the observed frame.  Due to
redshift, we expect to search for the presence of this line at
3.1~keV.  Jointly fitting the spectra
of three simultaneous observations of 3C 287 with {\it XMM-Newton}'s
EPIC/pn, MOS-1, and MOS-2 appears to exclude a very strong Fe K emission
line.  We investigate the possibility of a
narrow to broad emission line within the rest frame energy range of
6.4--6.97~keV.  Freezing the width of a Gaussian profile to zero in the
fit placed 

\centerline{~\psfig{file=f3.ps,width=3.3in,angle=-90}~}
\figcaption[h]{\footnotesize The {\it XMM-Newton} X-ray spectra of 3C
287 are shown above (black: pn, red: MOS1, blue: MOS2).  Energy units
are shown in the observed frame.  The spectra were fitted with a
simple absorbed power-law model.  The spectra themselves and the
data/model ratio indicate that no strong emission or absorption lines
are present.}
\medskip

\noindent a narrow line upper limit of 69 eV on the
equivalent width.  We find a broad line upper limit of 312 eV using the 
XSPEC model "diskline" instead of fitting a broad Gaussian line profile to 
account for the possibility of a relativistically broadened disk line.
However, the limits on the narrow line, which is expected to come
from a large region, could indicate that the innermost accretion flow may not be observed
directly.  This is consistent with the compact size of the radio emitting
region and supports the conclusion that a relativistically broadened iron
line is not present.  The limits on narrow line emission are typical for
Seyfert-1 AGN (Nandra et al. 1997).  The limits are less constraining when viewed in the context of high luminosity objects (see Bianchi et al.\ 2007), though there is considerable scatter in the strength of iron emission lines in such sources.

\subsection{OM Analysis}

To obtain a preliminary characterization of the broad-band spectral
energy distribution, we focused on a single characteristic OM
exposure, frame 433.  The OM V filter has a central wavelength of
5430\AA~ and spans roughly 1000\AA.  The task "omichain" generates a
list of source detections; 3C 287 was detected at the 12$\sigma$ level
of confidence, as imaged in Figure 2.  A count rate of 
$0.7\pm 0.1$ counts/second was recorded.  The measured instrumental 
magnitude is $V = 18.2\pm 0.2$.  Using the recipe on the SAS website 
appropriate for blue continuum sources, the count rate given above can 
be converted to a flux of $1.0 \pm 0.1 \times 10^{-12}~{\rm erg}~
{\rm cm}^{-2}~ {\rm s}^{-1}$. 

\subsection{Broad-band Analysis}

To place the X-ray flux trends from 3C 287 in context, we constructed
a broad-band spectral energy distribution using XSPEC.  Using the
ftool ``flx2xsp'' we generated single-bin spectral files.  To make use
of this tool, the flux in the bin is required, as well as the bin
boundaries.  The OM V filter spans approximately 1000\AA.  The 2MASS
survey recorded J and H magnitudes of 16.44 and 16.06, respectively.
These magnitudes were converted into fluxes using the Spitzer Science
Center conversion tool, and then into spectral files using the filter
boundaries appropriate to the 2MASS J and H filters (Cutri et al.\
2003).  VLA observations of 3C 287 recorded a flux density of 3.22 Jy
at 5 GHz (Kellermann et al.\ 1998).  We also converted this flux into
a spectral file.  For visual clarity, fiducial flux errors of 10\%
were assumed for the J, H, and radio fluxes, though true statistical
errors are far lower.

The resulting broad-band spectral energy distribution of 3C 287 is
shown in Figure 4.  The emission of 3C 287 peaks in the H, J, and UV
bands in the observed frame, or the optical and UV bands in the source
frame.  Thus, we do observe the ''Big Blue Bump'' emission that is
usually seen in radio-quiet quasars and interpreted as thermal emission
from an accretion disk (Shields 1978).  The OM measured luminosity is
about 6$\times 10^{45}$~ergs~s$^{-1}$ and it indicates high accretion
rates in this system.

The radio to X-ray luminosity ratio is smaller than typically observed
in samples of radio loud sources (blazars, Fossati et al. 1998; FRI galaxies,  Donato et al.\ 2004; FRII
type radio loud quasars and galaxies, Evans et al.\ 2006, Belsole et al.\ 2006, Balmaverde et al.\ 2006) in terms of the source being more powerful in radio than a
typical radio loud quasar (Elvis et al. 1994, Richards et al. 2006).
However, the ratio agrees with the ratios observed in compact radio
sources (GPS galaxies Guainazzi et al. 2006, Vink et al. 2006, GPS/CSS
quasars Siemiginowska et al. 2008).  In the case of GPS quasars the
X-ray emission could originate in a jet via Inverse Compton scattering
of the IR photons on the relativistic electrons in a parsec scale jet
(Blazejowski et al. 2004).  This emission could 
contribute to the observed X-ray emission. 

We also calculate the $\alpha_{ox}$ parameter (Avni \& Tannanbaum
1982) usually used to characterize a relative stregth of the
optical-UV thermal disk emission and the Comptonized X-ray component
(for example Sobolewska et al 2004).  This parameter is defined as:
$\alpha_{ox} = {\rm log [Flux(2500\AA)/Flux(2keV)]}/2.605$ in the
source rest frame.  We take the V-band magnitude and assume a spectral
slope of $f_{\nu} = {\nu}^{-0.5}$ to calculate the flux at the rest
frame $\lambda$=2500\AA~and the measured XMM flux at 2~keV.  Our
$\alpha_{ox}$ is equal to 1.7.  It is in a high end of a typical
distribution of the $\alpha_{ox}$ parameter for radio quiet quasars
($\alpha_{ox} = 1.5 \pm 0.2$, Kelly et al. 2008) and also radio loud quasars (Bechtold et al. 1994).

\section{Discussion and Conclusions}

In the sections above, we have detailed our analysis of a deep
exposure of 3C 287 made with {\it XMM-Newton}.  In the 0.3--10.0~keV
observed frame (approximately 0.6--20~keV in the source frame), the
spectrum is well represented by a simple absorbed power-law model.  In
AGN that are viewed at a modest inclination, it is common to observe
either an X-ray warm absorber (likely a disk wind; see, e.g., Reynolds
1997) or a soft excess (perhaps also tied to the disk; e.g. Crummy et
al.\ 2006), and an Fe K emission line due to hard X-ray illumination of
the disk, the torus, or both.  Fits made with a broken power-law were
made in order to test for evidence of either a soft excess or warm
absorption at low energy.  However, these fits do not provide a
significantly improved description of the data.  Similarly, fits with
Gaussian and disk line functions place limits on the strength of both
narrow and broad emission lines.

It is also worth noting that our results place a limit of
approximately $5 \times 10^{19}$~atoms~${\rm cm}^{-2}$ on any neutral
absorbing gas along our line of sight to 3C 287.  This limit is about
a factor of two lower than the measurement reported by Siemiginowska
et al.\ (2003).  Although models where the source is confined by dense gas (Snellen et al.\ 2000, Alexander 2000) have not been completely ruled out, new work argues against confinement (Guainazzi et al.\ 2004, Morganti 2007). 

Our X-ray results are consistent with emission related to a jet that
dominates over the typical X-ray emission associated 

\centerline{~\psfig{file=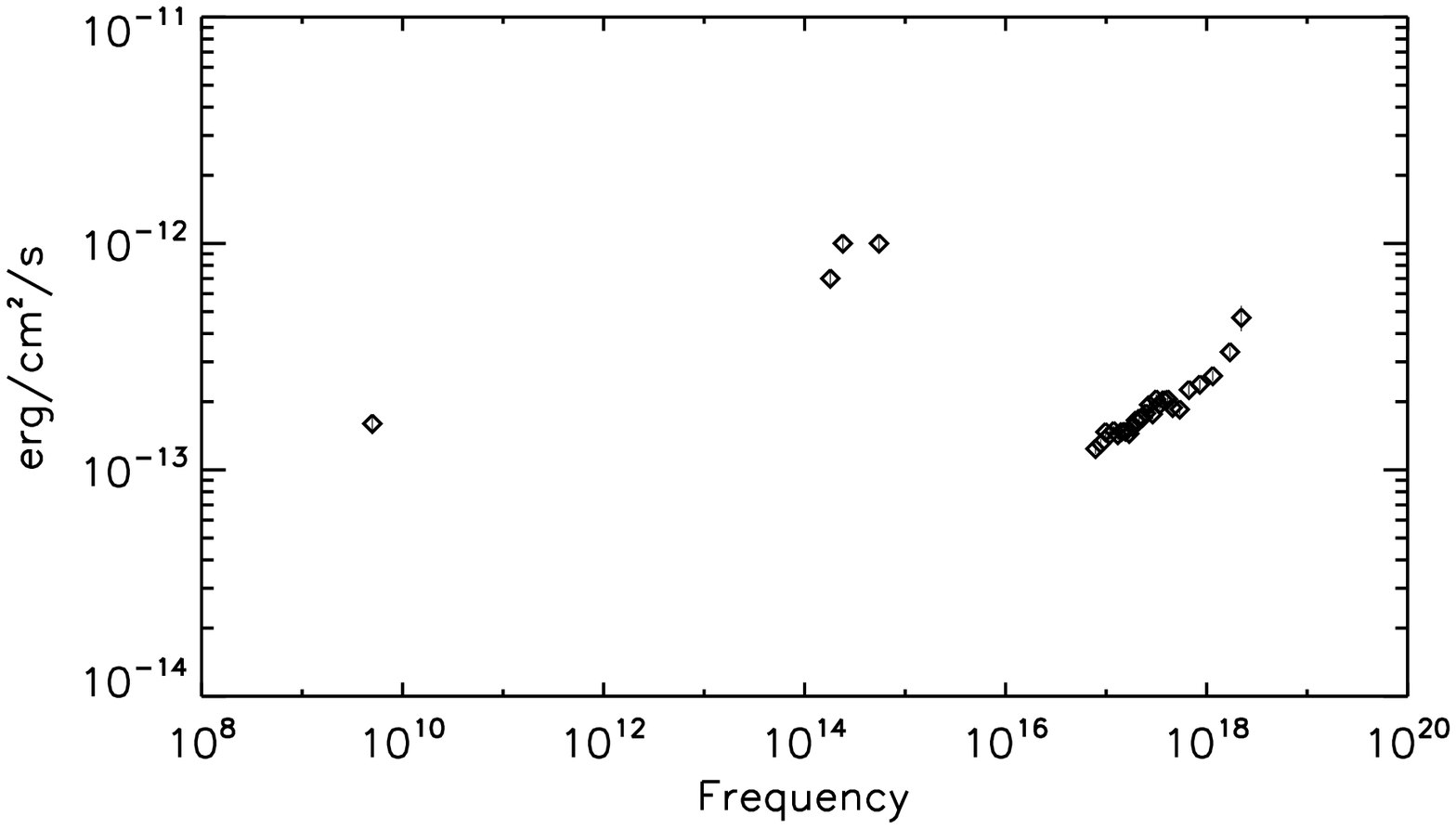,width=3.3in}~}
\figcaption[h]{\footnotesize A broad-band spectral energy distribution
of 3C 287 is shown above.  X-ray and optical points are from our {\it
XMM-Newton} observations.  Infrared points are from 2MASS observations
in J and H.  The single radio point is from a VLA observation at
5~GHz.}
\medskip

\noindent with the
accretion process.  The emission we observe could be accounted for
through a combination of synchrotron and synchrotron
self-Comptonization in the jet, hot spots, and compact lobes.

A line of sight coincident with the jet is suggested by deep
milliarcsecond radio imaging of 3C 287.  VLBA imaging at 15~GHz
reveals an extremely compact core in 3C 287, with a minimal extension
of less than 4 milliarcseconds, or less than about 20~pc in the source
frame (Kellermann et al.\ 1998).  This core is among the smallest and
least extended in the survey of AGN conducted by Kellerman et al.\
(1998), and is strongly suggestive of a jet that lies primarily along
our line of sight.  In many AGN, the presence of a strong,
narrow iron fluorescence line is tied to X-ray irradiation of a
parsec-scale torus (e.g. Antonucci 1993, Urry \& Padovani 1995, Yaqoob
\& Padmanabhan 2004, Nandra 2006).  Thus, although the radio core is
remarkably small by the standards of jet sources, it is still large
enough to block our line of sight to the innermost region (or even the torus) in 3C 287.
A possible difficulty with this inpretation is that emission consistent with the big blue bump is observed. This may indicate that the jet does not block all of the disk emission along our line of sight.

The question arises, then, why extreme blazar-like variability is not
observed in 3C 287 (and similar sources).  A partial 
answer may lie in the likely nature of CSS and GPS sources.
Proper motion studies of some sources in this class suggest that these
AGN have only been powering jets for approximatly $10^{3}$ years
(Conway 2002), however, a young radio source may have higher
efficiency of converting the jet power into radiation.  In such young
sources the X-ray emission may originate in compact lobes and hot
spots (Stawarz et al 2007) if the line of sight is not aligned
directly with a jet.  In this model there is no expectation of the
X-ray variability typically observed in blazars. Note that there is no
change in X-ray luminosity between the {\it Chandra} and {\it
XMM-Newton} observations (Siemiginowska et al 2008).

The broad-band spectral energy distribution  shown in Figure 4 may further
support this interpretation.  The source is more radio-loud than typical
radio-loud quasars (Elvis et al 1994, Richards, 2006, Siemiginowska et
al 2008), while its X-ray emission is relatively weak with
$\alpha_{ox}=1.7$. This suggests a weak contribution from the standard
emission component (for example a parsec scale jet, Blazejowski et al.
2004) observed in radio-loud quasars.

We conclude that 3C 287 may be a ``proto-blazar'' candidate.
Dedicated multi-wavelength monitoring of sources like 3C 287 over a
period of years can help to better reveal the nature of this source.

\vspace{0.25in}

We thank the anonymous referee for insightful comments that improved the paper.  JMM gratefully acknowledges funding from NASA, through the {\it
XMM-Newton} guest observerations program.  Partial support for
this work was provided by the National Aeronautics and Space
Administration through Chandra Award Number GO2-3148A and GO5-6113X
and under the contract NAS8-39073.

\begin{table}[ht]
\caption~\center{Spectral Fit Parameters}
\begin{center}				
\begin{footnotesize}
\begin{tabular}{l c}			
\hline
\hline \\				
Model Parameter & EPIC/pn \\ [0.5 ex] 
\hline \\				

Power-Law Model \\
\hline
N$_{H}$ ($10^{19}$ cm$^{-2}$)	& $<$ 3.1		\\
$\Gamma$				& 1.72 $\pm$ 0.02	\\
Norm. ($10^{-4}$~ph~cm$^{-2}$~s$^{-1}$)	& 1.04 $\pm$ 0.02	\\
$\chi^{2}/\nu$				& $650.3/598 = 1.09$	\\ [1 ex]

Flux ($10^{-13}$ ergs cm$^{-2}$ s$^{-1}$)	& 7.1 $\pm$ 0.2	\\	
Lumin ($10^{45}$ ergs s$^{-1}$)		& 5.8 $\pm$ 0.2	\\ [1 ex]

\hline \\

Broken Power-Law Model \\
\hline
N$_{H}$ ($10^{19}$ cm$^{-2}$)		& $<$ 18 \\
$\Gamma_{1}$					& 1.74 $\pm$ 0.09 \\
$\Gamma_{2}$					& 1.5 $\pm$ 0.2 \\
E$_{break}$ (keV)				& 4.1 $\pm$ 0.3 \\
Norm. ($10^{-4}$~ph~cm$^{-2}$~s$^{-1}$)		& 1.04 $\pm$ 0.06 \\
$\chi^{2}/\nu$					& $643.4/596 = 1.08$ \\ [1 ex]

Flux ($10^{-13}$ ergs cm$^{-2}$ s$^{-1}$)	& 7.3 $\pm$ 0.5	\\
Lumin ($10^{45}$ ergs s$^{-1}$)			& 5.9 $\pm$ 0.5	\\ [1 ex]

\hline \\

Line Profile Fits to Power-Law Model \\
(Narrow Line) : (Broad Line) \\
\hline
N$_{H}$ ($10^{19}$ cm$^{-2}$)	& $<$ 3.5 : $<$ 3.7 \\
$\Gamma$			& 1.73 $\pm$ 0.02 : 1.72 $\pm$ 0.03 \\
Norm.$_{Power-Law}$ ($10^{-4}$~ph~cm$^{-2}$~s$^{-1}$)
				& 1.04 $\pm$ 0.02 : 1.04 $\pm$ 0.02 \\
E$_{line}$ (keV)		& 3.2 : 3.1   \\
Norm.$_{Line}$ ($10^{-6}$~ph~cm$^{-2}$~s$^{-1}$) & $<$ 1.0 : $<$ 4.3  \\
$\chi^{2}/\nu$		& $644.9/594 = 1.09$ : $645.5/592 = 1.09$ \\ [1 ex]

Flux ($10^{-13}$ ergs cm$^{-2}$ s$^{-1}$)  & 7.1 $\pm$ 0.2 : 7.1 $\pm$ 0.3 \\
Lumin ($10^{45}$ ergs s$^{-1}$)		& 5.8 $\pm$ 0.1 : 5.7 $\pm$ 0.3 \\
EW (eV)					& 69 : 312	\\ [1 ex]

\hline
\end{tabular}

\tablecomments{The results of joint spectral fits to the EPIC pn,
MOS1, and MOS2 spectra of 3C 287 are detailed above.  Though a broken
power-law provides a slightly better fit, the spectrum is consistent
with a simple power-law form.  The narrow line is fit with a Gaussian model
while the broad line is fit with a diskline model.  Upper limits are given
for the column density of neutral gas along the line of sight and for line
profile normalizations.  The bolometric luminosity is quoted in the table
assuming that X-rays contribute 10$\%$ of the bolometric luminosity.  We
report the best-fit X-ray flux values over the 0.3--10.0 keV band.
Reported flux and luminosity of 3C 287 in the text are taken from the
power-law fits.}
\vspace{-1.0\baselineskip}
\end{footnotesize}
\end{center}
\end{table}


\begin{references}

\reference{} Alexander, P., 2000, MNRAS, 319, 8

\reference{} Antonucci, R., 1993, ARA\&A, 31, 473

\reference{} Arnaud, K. A., 1996, ADASS V, eds. G. Jacoby and
J. Barnes, ASP Conf.\ Series 101, 17

\reference{} Avni, Y., \& Tananbaum, H.\ 1982, ApJ.Lett, 262, L17 

\reference{} Balmaverde, B., Capetti, A., \& Grandi, P., 2006, A\&A, 451, 35

\reference{} Bechtold, J., et al.\ 1994, AJ, 108, 374 

\reference{} Belsole, E., Worrall, D. M., \& Hardcastle, M. J., 2006, MNRAS, 366, 339

\reference{} Bianchi, S., Guainazzi, M., Matt, G., \& Fonseca Bonilla, N., 2007, A\&A, 467, L19

\reference{} B{\l}a{\.z}ejowski, M., Siemiginowska, A., Sikora, M., Moderski, R., 
\& Bechtold, J.\ 2004, ApJLett, 600, L27 

\reference{} Conway, J. E., 2002, NewAR, 46, 263

\reference{} Crummy, J., Fabian, A. C., Gallo, L., \& Ross, R. R.,
2006, MNRAS, 365, 1067

\reference{} Cutri, R. M., et al., 2003, ``The IRSA 2MASS All-Sky
Survey Point Source Catalog'', NASA/IPAC Infrared Science Archive

\reference{} Donato, D., Sambruna, R. M., \& Gliozzi, M., 2004, ApJ, 617, 915

\reference{} Elvis, M., et al.\ 1994, ApJSupp, 95, 1 

\reference{} Evans, D. A., et al., 2006, ApJ, 653, 1121

\reference{} Fender, R. P., 2004, ``Jets from X-ray Binaries'', to
appear in ``Compact Stellar X-ray Sources'' eds. W. H. G. Lewin \&
M. van der Klis, Cambridge, 2004, see astro-ph/0303339

Gallo, E., Fender, R., \& Pooley, G. G., 2003, MNRAS, 344, 60

\reference{} Fossati et al.(1998){1998MNRAS.299..433F} Fossati, G., Maraschi,
L., Celotti, A., Comastri, A., \& Ghisellini, G.\ 1998, \mnras, 299, 433

\reference{} George, I. M., \& Fabian, A. C., 1991, MNRAS, 249, 352

\reference{} Guainazzi, M., Siemiginowska, A., Rodriguez-Pascual, P., \& Sgnhellini, C., 2004, A\&A, 421, 461

\reference{} Guainazzi et al.(2006){2006A\&A...446...87G} Guainazzi, M.,
Siemiginowska, A., Stanghellini, C., Grandi, P., Piconcelli, E.,
\& Azubike Ugwoke, C.\ 2006, \aap, 446, 87

\reference{} Kellermann, K. I., et al., 1998, ApJ, 115, 1295

\reference{} Kelly, B.~C., Bechtold,  J., Trump, J.~R., Vestergaard, M., 
\& Siemiginowska, A.\ 2008, ApJ in press, see  ArXiv e-prints, 801, arXiv:0801.2383 

\reference{} Laor, A., 1991, ApJ, 376, 90

\reference{} Miller, J. M., 2007, ARA\&A, 45, in press, arxiv:0705.0540

\reference{} Morganti, R., 2007, arxiv:0710:1197

\reference{} Nandra, K., George, I. M., Mushotzky, R. F., Turner, T. J., \&
Yaqoob, T., 1997, ApJ, 477, 602

\reference{} Nandra, K., 2006, MNRAS, 368, L62

\reference{} Reynolds, C. S., 1997, MNRAS, 286, 513

\reference{} Richards, G.~T., et  al.\ 2006, ApJSupp, 166, 470 

\reference{} Shields, G.~A.\ 1978, Nature,  272, 706 

\reference{} Siemiginowska, A. et al 2008, ApJ. submitted

\reference{} Siemiginowska et al.(2008){2008arXiv0804.1564S} Siemiginowska,
A., LaMassa, S., Aldcroft, T.~L., Bechtold, J., \& Elvis, M.\ 2008,
ArXiv e-prints, 804, arXiv:0804.1564

\reference{} Siemiginowska, A., et al., 2003, PASA, 20, 113

\reference{} Snellen, I. A. G.,Schilizzi, R. T., Miley, G. K., de Bruyn, A. G., Bremer, M. N., \& Rottgering, H. J. A., 2000, MNRAS, 319, 445

\reference{} Sobolewska, M.~A., Siemiginowska, A., {\. Z}ycki, P.~T.\ 2004, ApJ, 
608, 80

\reference{} Stawarz, L., Ostorero, L., Begelman, M.~C., Moderski, R., Kataoka, J., 
\& Wagner, S.\ 2007, ArXiv e-prints, 712, arXiv:0712.1220 

\reference{} Urry, C. M., \& Padovani, P., 1995, PASP, 107, 803

\reference{} Vink et al.(2006){2006MNRAS.367..928V} Vink, J., Snellen, I.,
Mack, K.-H., \& Schilizzi, R.\ 2006, \mnras, 367, 928 

\reference{} Yaqoob., T., \& Padmanabhan, U., 2004, ApJ, 604, 63

\end{references}
\end{document}